\pdfoutput=1
\documentclass[12pt,reqno]{article}
\usepackage{jheppub}
\usepackage{amsmath,amssymb,amsfonts}
\graphicspath{ {./Graphs/} }
\usepackage{float}             
 \usepackage[T1]{fontenc}
\usepackage[utf8]{inputenc}
\usepackage{mathtools}  
\usepackage{nameref}
\usepackage{blkarray}
\usepackage[usenames,dvipsnames]{xcolor}
\usepackage{epsfig}
\usepackage{epstopdf}
\usepackage{dcolumn}
\usepackage{tikz}
\usetikzlibrary{shapes.geometric, arrows}
\usepackage{upgreek}
\usepackage{setspace}
\usepackage{enumitem}
\usepackage{array,multirow,bigdelim}
\usepackage{cleveref}
\usepackage{bm}

\def\be{\begin{equation}}
\def\ee{\end{equation}}
\def\ba{\begin{eqnarray}}
\def\ea{\end{eqnarray}}

\newcommand{\bea}{\begin{eqnarray}}
\newcommand{\eea}{\end{eqnarray}}
\newcommand{\bean}{\begin{eqnarray*}}
\newcommand{\eean}{\end{eqnarray*}}

\renewcommand{\tilde}{\widetilde}

\renewcommand{\phi}{\varphi}

\renewcommand{\phi}{\varphi}

\preprint{
SAGEX-19-24-E}
\title{Post-Minkowskian Scattering Angle in Einstein Gravity} 
\author{N.~E.~J.~Bjerrum-Bohr,}
\author{Andrea~Cristofoli,}
\author{Poul~H.~Damgaard}
\affiliation{Niels Bohr International Academy and Discovery Center\\ 
The Niels Bohr Institute, University of Copenhagen\\
Blegdamsvej 17, DK-2100 Copenhagen, Denmark}
\emailAdd{bjbohr@nbi.dk}
\emailAdd{a.cristofoli@nbi.ku.dk}
\emailAdd{phdamg@nbi.dk}
\date{\today}
\abstract{Using the implicit function theorem we demonstrate that solutions to the classical part of the relativistic Lippmann-Schwinger equation are in one-to-one correspondence with those
of the energy equation of a relativistic two-body system. A corollary is that the scattering angle can be
computed from the amplitude itself, without having to introduce a potential. All results are universal and provide for the case of general relativity a very simple formula for the scattering angle in terms of the classical part of the amplitude, to any order in the post-Minkowskian expansion.}
\begin{document}
\maketitle
\section{Introduction}\label{sec:introduction}
The Post-Minkowskian expansion of general relativity promises to become a new and powerful tool with which to compute observables of two-body gravitational
interactions \cite{Damouruno,Damourdue,Bjerrum-Bohruno,Cheung,Bernuno,Antonelli,Cristofoliuno,Berndue,Kalin}. As a systematic expansion in Newton's
constant $G_N$, the Post-Minkowskian framework is perfectly suited for a standard second-quantized field theory approach to classical gravity \cite{Damourdue,Bjerrum-Bohruno,Cheung}. There is now hope that modern field theory techniques may radically change the prospect for how far analytical calculations can be pushed
in general relativity. Currently also much work goes into seeing how Post-Minkowskian gravitational interactions of classically spinning objects can be treated
by modern quantum field theory techniques \cite{Vinesuno,Vinesdue,Guevarauno,Chung,Maybeeuno,Guevaradue,Nima,Damgaard}, leading again to a complete
revision of how such classical observables can be computed in general relativity.\\[5pt]
When the Post-Minkowskian expansion is applied to the two-body bound-state problem it is natural to phrase it in terms of a potential $V$, either as provided implicitly
through the Effective One-Body Hamiltonian \cite{Buonanno} or by the large-distance effective Hamiltonian obtained by matching of amplitudes \cite{Cheung}. Up to
canonical transformations, this is equivalent to studying the relativistic Salpeter equation \cite{Cristofoliuno} based on an Hamiltonian operator\\[-10pt]
\begin{equation}
\label{opeH}
\hat{\mathcal{H}}\;=\; \hat{\mathcal{H}_0}+ \hat{V} \;=\; \sum_{i=1}^{2}\sqrt{\hat{p}^2+m^2_i}+\hat{V}\,,
\end{equation}
and then taking the classical limit. Only the positive-energy solutions enter in this Hamiltonian because we remove
antiparticles in the scattering process by hand
when taking the macroscopic classical limit. The momentum-space potential $\tilde{V}$ can be easily computed by solving
the associated Lippmann-Schwinger equation for the full scattering amplitude \cite{Cristofoliuno},\\[-10pt]
\begin{equation}
\mathcal{M}(p,p')\;=\;\tilde{V}(p,p')\;+\;\int \frac{d^{3}k}{(2 \pi)^3}\frac{\tilde{V}(k,p)\;\mathcal{M}(k,p')}{E_p-E_k+i \epsilon}\,,
\end{equation}\\[-10pt]
inverting it,\\[-10pt]
\begin{equation}
\tilde{V}(p,p') \;=\; \mathcal{M}(p,p') - \int\frac{d^{3}k}{(2\pi)^3}
\frac{\mathcal{M}(p,k)\;\mathcal{M}(k,p')}{E_p - E_k + i\epsilon} + \ldots\,,
\end{equation}
and taking the classical limit. This is the systematics of the {\em Born subtractions} needed to define a potential from the scattering amplitude.  
The so-called super-classical terms \cite{Kosower} cancel in the process, rendering the classical limit of the potential well-defined.
By performing a suitable Fourier transform, this leads to the conventionally defined position-space potential $V$ in the chosen coordinates. \\[4pt]
It should be noted that the effective field theory matching employed in refs. \cite{Cheung,Bernuno,Berndue} is equivalent to the method of Born subtractions
\cite{Cristofoliuno}. In four dimensions, the effective field theory matching, after suitable reduction of the four-dimensional amplitude integrals to integrals living in only three dimensions, involves cancellations of identical integrals, which hence do not need to be evaluated. The same cancellations among three-dimensions integrals
can be achieved also in the Born subtraction method (indeed, the two methods are completely equivalent), but we prefer to evaluate all integrals for clarity. \\[4pt]
The position-space potential $V$ seems needed when solving the bound-state problem in general relativity. However, this quantity is not very natural in the field theoretic framework where everything is based on the gauge invariant $S$-matrix with incoming and outgoing momenta defined at
Minkowskian infinity. One would surely prefer as far as possible a formulation in which $V$ would not be needed. This problem is compounded when we consider
a coordinate-independent observable such as the classical scattering angle from far infinity to far infinity. Conventionally, we will be led to solve the classical
analog of the Salpeter Hamiltonian of eq. (\ref{opeH}) and then follow the classical analysis of the scattering
problem \cite{Buonanno}. While that method is correct, it seems intuitively surprising that it should be necessary to go through the carefully Born subtracted position-space
potential $V$ as an intermediate step. Indeed, we know from quantum field theory that all scattering information from far infinity to far infinity is contained in
the  $S$-matrix, viz., the scattering amplitude.\\[4pt]
This puzzle has become greatly clarified by the observation of Bern et al. \cite{Berndue} that up to two loop order (3PM order in the Post-Minkowskian
counting) seemingly miraculous cancellations take place, leaving a perturbatively expanded expression for the two-loop scattering angle expressed entirely in
terms of the classical part of the scattering amplitude up to that two-loop order. If this phenomenon is to persist to all orders it means that all the classical pieces of the
Born subtractions defined above provide a potential $V$ in precisely such a manner as to compensate, exactly and to all orders in the
coupling $G_N$, the additional terms that arise from solving the expanded classical Salpeter equation. While the apparent conspiracy of two such totally unrelated
equations having a one-to-one relation might seem improbable, we shall in this paper elucidate how this indeed will be true. Our tool will be the implicit function
theorem that sometimes goes under the name of Dini's Theorem (although a different theorem also carries Dini's name). In the process we will unravel new
and compact relations between the classical potential, together with its derivatives, and the classical part of the scattering amplitude.\\[4pt]
Having this relationship established, a next burning question is: how do we then compactly express the scattering angle directly in terms of the classical
part of the scattering amplitude? To find such an expression, we make use of an idea proposed by Damour in ref. \cite{Damourdue}, mapping the classical and
fully relativistic Salpeter Hamiltonian into an auxiliary Hamiltonian that is formally in the non-relativistic form of a one-particle Hamiltonian for a particle of mass equal to 
1/2 in appropriate units and with a potential that is only position-dependent. Considering the quantized analog of this Hamiltonian one immediately proves, in 
essentially one line, that the solution for the scattering angle indeed only depends on the classical part of the scattering amplitude.\footnote{While this paper was in preparation,
the same observation was made in ref. \cite{Kalin}.} But armed with the non-relativistic auxiliary problem we can do far more than that. Indeed, the classical part
of the mapped scattering problem must now be WKB-exact and even solved by only its leading-order piece of order $\hbar^0$ in the exponent. This is, consistently,
simply the classical Hamilton-Jacobi equation with the phase identified with the generating function $S$. Much literature exists on the relationship between the scattering angle and the WKB-approximation as well as their 
relation to the eikonal limit, and we hope our discussion
here will clarify some confusion. Our end result is a very simple formula for the scattering angle in terms of the classical part of the scattering amplitude, to all orders in the coupling.
\section{The Lippmann-Schwinger equation in position space}
The Lippmann-Schwinger equation is usually expressed as an integral equation involving amplitudes and potentials in momentum space. For the case of non-relativistic systems, its space representation states that the Fourier transform of the classical part of the amplitude is proportional to the potential. However, for the case of fully relativistic systems, this is no longer true \cite{Cristofoliuno}. We shall here extend this observation by demonstrating that the position-space representation of the Lippmann-Schwinger equation for fully relativistic systems can be expressed as a differential equation for the potential and the classical part of the amplitude. To show this, we start by considering the fully relativistic Lippmann-Schwinger equation in momentum space \\[-7pt]
\begin{equation}
\label{lip}
\mathcal{M}(p,p')\;=\;\tilde{V}(p,p')+\int \frac{d^3k}{(2 \pi)^3}\frac{\tilde{V}(k,p)\,\mathcal{M}(k,p')}{E_p-E_k+i \epsilon}\,.
\end{equation}
Kinematics will always be that of the center of mass frame. We parametrize the potential in momentum space as
\cite{Berndue}\\[-7pt]
\begin{equation}
\label{pot}\displaystyle
\!\!\tilde{V}(k_i,k_j)\;=\;\sum_{n=1}^{\infty}\bigg(\!\frac{G_N}{2}\!\bigg)^n(4\pi)^{\frac{3}{2}}\frac{\Gamma(\frac{3-n}{2})}{\Gamma(\frac{n}{2})}\frac{c_n(k_i,k_j)}{|k_i-k_j|^{3-n}}\,,\ \ \  c_{n}(k_i,k_j)\;=\;c_n\bigg(\frac{k^2_i+k^2_j}{2}\bigg)\,.
\end{equation}
Eq. (\ref{lip}) allows us to express the momentum-space amplitude as \\[-7pt]
\begin{equation}\begin{split}
\label{schw}
\mathcal{M}(p,p')&\!=\!\sum_{n=0}^{\infty}\int_{k_1,k_2,\ldots,k_n}\!\!\frac{\tilde{V}(p,k_1)\,\tilde{V}(k_1,k_2)\,\cdots\, \tilde{V}(k_n,p')}{(E_p\!-\!E_{k_1})(E_{k_1}\!-\!E_{k_2})\cdots(E_{k_{n-1}}\!-\!E_{k_n})}\!=\!\tilde{V}(p,p')\!+\!\sum_{n=1}^{\infty}\!S_{n}(p,p')\,,\\[-5pt]
\end{split}\end{equation}\\[-10pt]
where the $n$-th terms of the series has $n+1$ factors of potential $\tilde{V}$ in the numerator and $n$ energy denominators. \\[3pt]
We are only interested in the classical pieces of this equation, which means that we must device a precise
mechanism to discard super-classical and quantum terms from the right hand side of eq. (\ref{schw}) based on the
$\hbar$-counting \cite{Kosower}. In order to understand this procedure, we start by considering the first non-trivial ($n=1$) 
term of eq. (\ref{schw}) and then extend the reasoning to all $n$. 
For $n=1$ we have\\[-5pt]
\begin{equation}
\label{ptiu}
S_1\;=\;\int \frac{d^3k}{(2\pi)^3}\frac{\tilde{V}(p,k)\,\tilde{V}(k,p')}{E_p-E_k}\,.
\end{equation}
Since we are only interested in classical terms, we can expand the propagator in eq. (\ref{ptiu}) around $k^{2}_{i}=k^{2}_{j}$ as\\[-6pt]
\begin{equation}
\label{propa}
\frac{1}{E_{k_{i}}-E_{k_j}}\;=\;\frac{2E \xi}{k^2_i-k^2_j}+\frac{3\, \xi-1}{2 E \xi}+\ldots\,, \quad \xi\;\equiv\; \frac{E_a\,E_b}{E^2}\,, \quad E \;\equiv\; E_a+E_b\,.
\end{equation}
Using this expansion, the only classical contributions that could arise from (\ref{ptiu}) are\\[-8pt]
\begin{equation}
S_1\;=\;2 E \xi I_1+\bigg(\frac{3\xi-1}{2E\,\xi}\bigg)J_1+\ldots\,,
\end{equation}\\[-15pt]
where \\[-12pt]
\begin{equation}
I_1 \;\equiv\; \int \frac{d^3k}{(2\pi)^3}\frac{\tilde{V}(p,k)\,\tilde{V}(k,p')}{p^2-k^2} \ , \quad J_1 \;\equiv\; \int \frac{d^3k}{(2\pi)^3}\tilde{V}(p,k)\,\tilde{V}(k,p')\,.
\end{equation}\\[-5pt]
We start by evaluating the classical contributions from $I_1$, using (\ref{pot})\\[-5pt]
\begin{equation}
\label{i2}
I_1\;=\;(4\pi)^3\sum_{n,m=1}^{\infty}\bigg(\frac{G_N}{2} \bigg)^{n+m}\frac{\Gamma(\frac{3-n}{2})\,\Gamma(\frac{3-m}{2})}{\Gamma(\frac{n}{2})\,\Gamma(\frac{m}{2})}\int\frac{d^3k}{(2\pi)^3}\frac{c_n(p,k)\,c_m(k,p')}{(p^2-k^2)\,|k-p|^{3-n}\,|k-p'|^{3-m}}\,.
\end{equation}
In order to discard super-classical and quantum terms we expand the numerator around $k^2=p^2$ as\\[-5pt]
\begin{equation}
c_{n}(k,p)\,c_{m}(k,p')=c_n^0\,c^0_m+\frac{1}{2}(c^0_{n}\,\partial_{p^2}\,c^{0}_{m}+c^{0}_{m}\,\partial_{p^2}\,c^{0}_{n})(k^2-p^2)+\ldots\,, \quad c^0\;\equiv\; c_{|_{k^2=p^2}}\,.
\end{equation}
The $\hbar$-counting thus tells us that the only classical contribution ({\it cl.}) from eq. (\ref{i2}) is given by\\[-8pt]
\begin{equation}
I_{1}^{cl.}\;=\;-(4\pi)^3\sum_{n,m=1}^{\infty}\bigg(\frac{G_N}{2} \bigg)^{n+m}\frac{\Gamma(\frac{3-n}{2})\,\Gamma(\frac{3-m}{2})}{\Gamma(\frac{n}{2})\,\Gamma(\frac{m}{2})}\frac{(c^0_{n}\partial_{p^2}c^{0}_{m}+c^{0}_{m}\partial_{p^2}c^{0}_{n})}{2}\,G^{(2)}_{n,m}(q)\,,
\label{i2cla}
\end{equation}
where we have introduced $q \equiv p' - p$ and\\[-5pt]
\begin{equation}
G^{(2)}_{n,m}(q) ~\equiv~ \int\frac{d^3k}{(2\pi)^3}\frac{1}{|k|^{3-n}\,|k-q|^{3-m}}\,.
\end{equation}
It is also convenient to define its Fourier transform\\[-5pt]
\begin{equation}
g^{(2)}_{n,m}(r) 
\;\equiv\; \int \frac{d^3q}{(2\pi)^3}G^{(2)}_{n,m}(q)e^{i q \cdot r }\,,
\end{equation}
which is seen to factorize,\\[-10pt]
\begin{eqnarray}
g^{(2)}_{n,m}(r) &=& \int\!\!\!\frac{d^3k}{(2\pi)^3}\int\!\!\! \frac{d^3q}{(2\pi)^3}\frac{e^{i (q+k) \cdot r}}{|k|^{3-n}\,|q|^{3-m}}
= \int\!\!\frac{d^3q}{(2\pi)^3}\frac{e^{i q \cdot r}}{|q|^{3-m}}\times\int\!\!\frac{d^3k}{(2\pi)^3}\frac{e^{i k \cdot r}}{|k|^{3-n}}=g_n(r)\,g_m(r)\nonumber\,.\\ && 
\end{eqnarray}\\[-25pt]
The function $g_n(r)$ is well known and given by\\[-5pt]
\begin{equation}
g_n(r)\; =\; \frac{\Gamma(\frac{n}{2})}{\Gamma(\frac{3-n}{2})}\bigg(\frac{2}{r}\bigg)^n\frac{1}{(4\pi)^{\frac{3}{2}}}\,.
\end{equation}
Using this, the position-space representation of eq. (\ref{i2cla}) becomes\\[-5pt]
\begin{equation}
\tilde{I}_1^{cl.}\;=\;-\sum_{n,m=1}^{\infty}\bigg(\frac{G_N}{r}\bigg)^{n+m}(c^0_{n}\,\partial_{p^2}\,c^{0}_{m}) \,, \quad \tilde{I}_1^{cl.} \;\equiv\; \int\frac{d^3q}{(2\pi)^3}\,e^{i q \cdot r}
\,I_{1}^{cl.} ~.
\end{equation}
This can be expressed in an even simpler form by realizing that it can be factorized,\\[-5pt]
\begin{equation}
\label{xspace}
\tilde{I}_1^{cl.}\;=\;-\bigg[\sum_{n=1}^{\infty}\bigg(\frac{G_N}{r} \bigg)^n c^0_n\bigg]\bigg[\sum_{m=1}^{\infty}\bigg(\frac{G_N}{r} \bigg)^m \partial_{p^2}\,c^0_m \bigg]\,.
\end{equation}
This nicely connects with the Fourier transform of the potential in position space,\\[-10pt]
\begin{equation}
V(p,r)\;=\;\sum_{n=1}^{\infty}\bigg(\frac{G_N}{r} \bigg)^n c_n(p^2)\,,
\end{equation}\\[-10pt]
giving\\[-5pt]
\begin{equation}
\tilde{I}_1^{cl.} \;=\;- V(p,r)\,\partial_{p^2}V(p,r) ~.
\end{equation}
As for the remaining integral, one has \\[-5pt]
\begin{equation}
J_1\;=\;(4\pi)^3\sum_{n,m=1}^{\infty}\bigg(\frac{G_N}{2} \bigg)^{n+m}\,\frac{\Gamma(\frac{3-n}{2})\,\Gamma(\frac{3-m}{2})}{\Gamma(\frac{n}{2})\,\Gamma(\frac{m}{2})}\,c^0_{n}\,c^{0}_{m}G^{(2)}_{n,m}(q)\,.
\end{equation}
One readily finds that its Fourier transform $\tilde{J}_1^{cl.}$ simply satisfies $\tilde{J}_1^{cl.}=V^2$.
Defining the real-space representation of the classical part of the amplitude by\\[-5pt]
\begin{equation}
\tilde{\mathcal{M}}^{cl.}(p,r)\;\equiv\; \int \frac{d^3q}{(2\pi)^3}\,\mathcal{M}^{cl.}(p,p')\,e^{i q \cdot r}\,,
\end{equation}
we find that the leading first term to all orders in $G_N$ is given by\\[-5pt]
\begin{equation}
\label{state1}
\tilde{\mathcal{M}}^{cl.}(p,r)\;=\;V-2E\xi \,V \partial_{p^2}V+\bigg(\frac{3 \xi-1}{2E\xi} \bigg)\,V^2+\ldots\,.
\end{equation}
As for the remaining terms in the series, they can be evaluated in exactly the same fashion by an expansion of the energy denominators and numerators, although the complexity of these analytical expressions grow rapidly and we do not display them here. (Remarkably, the classical part of the series can always be expressed as a linear combination of generalized $n$-loop massless sunset diagrams with external momentum $q$; this is shown in the Appendix).
We have thus shown that a quite simple differential equation links the classical part of the amplitude to the potential. At higher loop
level the order of the differential equation increases, but the structure remains. Let us now interpret this relation by considering it order by order in powers of $G_N$. At linear order in $G_N$ the relation is trivial and simply states that the Fourier transform of the amplitude at tree level is the potential, a textbook observation. At quadratic order things become more interesting and one has\\[-5pt]
\begin{equation}
\mathcal{\tilde{M}}^{cl.}_{1-loop}(p,r)\;=\;V_{G_N^2}-2E\xi \,V_{G_N}\partial_{p^2}V_{G_N} +\bigg(\frac{3 \xi -1}{2E \xi} \bigg)\,V^2_{G_N}\,.
\end{equation} 
By using the definition of potential in position space one has\\[-5pt]
\begin{equation}
\label{fainait}
\mathcal{\tilde{M}}^{cl.}_{1-loop}(p,r)\;=\;\frac{G^2_N}{r^2}\bigg[c_2-2E\xi\, \partial_{p^2}c_1 \,c_1 +\bigg(\frac{3 \xi -1}{2E \xi} \bigg)\,c^2_1\bigg]\,.
\end{equation}
This relation reproduces exactly the classical part of the 2PM amplitude in position space\\[-10pt]
\begin{equation}
\mathcal{\tilde{M}}^{cl.}_{1-loop}(p,r)\;=\;\frac{3(m_1+m_2)\,(m^2_1\,m^2_2-5\,p_1 \cdot p_2^2)}{4E^2\xi}\frac{\,G^2_N}{r^2}\,.
\end{equation}
It is elementary, although tedious, to derive the analogous relations to any higher loop order and there is no need to reproduce those more complicated expressions here. What is far more interesting is the fact that precisely the same series can be understood also from an alternative point of view by applying the implicit function theorem to the relativistic energy equation.  

\section{Dini's theorem and the Lippmann-Schwinger equation}
We start by stating the implicit function theorem (Dini's theorem) in a form useful for the present purpose:
\vspace{2mm}
\newline
\emph{Let $F: \mathbb{R}^2 \rightarrow \mathbb{R}$ be a $\mathcal{C}^{\infty}$ function. Consider a point $(x_{0},y_{0})$ such that $F(x_0,y_0)=0$ and $\partial_{x}F(x_{0},y_{0}) \neq 0$. Then there exist a closed neighbourhood of $(x_{0},y_{0})$ and a function $y=f(x)$ so that $F(x,y(x)) =0$ for every point in that neighbourhood. The implicit function $y=f(x)$ will admit a Taylor expansion in terms of the partial derivatives of $F(x,y)$ given by\\[-25pt]
\begin{eqnarray}
y(x)\;&=&\;y(x_0)+y'(x_0)(x-x_0)+\frac{1}{2}y''(x_0)(x-x_0)^2+\ldots\,,\\
\label{der1}
\hskip-6.4cm y'(x_0)\;&=&\;-\left.\frac{\partial_{x}F}{\partial_{y}F}\right|_{x=x_0\:,\; y=y({x_{0}})}\,,\\
\label{der2}
\hskip-0.76cm y''(x_0)\;&=&\;-\left.\frac{\partial^{2}_{xx}F+2y' \partial^2_{xy}F +\partial^2_{yy}F y'^{2}}{\partial_{y}F}\right|_{x=x_0\:,\; y=y({x_{0}})}\,,
\end{eqnarray}
where the higher order derivatives can be computed from \\[-5pt]
\begin{equation}
\bigg(\partial_{x}+\frac{dy}{dx} \partial_{y}  \bigg)^{n}F(x,y(x))\;=\;0 \quad , \quad \forall n \in \mathbb{N}\,,
\end{equation}
by the binomial expansion of operators.
}
\vspace{2mm}
\newline
We now apply this theorem to the problem of inverting the relativistic energy equation in terms of three-momenta. This is precisely what arises in the post-Minkowskian two-to-two scattering process where we must solve the classical energy relation of eq. (\ref{opeH}),\vskip-0.3cm
\begin{equation}
\label{cons}
\sum_{i=1}^{2}\sqrt{p^2+m^2_i}+V(p,r)\;=\;E \ , \quad 
V(p,r)\;=\;\sum_{n=1}^{\infty} \bigg(\frac{G_{N}}{r}\bigg)^n c_{n}(p^2)\,.
\end{equation}
In order to find a solution to eq. (\ref{cons}) we apply Dini's theorem by choosing $p^2$ as $y$ and $G_{N}$ as $x$ respectively.\footnote{We choose $G_N$ for sheer convenience because post-Minkowskian Hamiltonians in the center of mass frame have the same counting in $1/r$ and $G_N$. In case of higher-derivative gravity this counting is of course broken by new coupling constants \cite{Gab,Emond,Cristofolidue}. That more general case can be analyzed analogously by simply identifying $y$ with $r$.}  Then,\\[-25pt]
\begin{eqnarray}
F(p^2,G_N)\;&=&\;\sum_{i=1}^{2}\sqrt{p^2+m^2_i}+V(p,r)-E \,,\\
F(p^2(G_N),G_N)\;&=&\;0 \ , \quad \partial_{G_N}F(p^2,G_N)\;=\;\partial_{G_N}V \neq 0\,.
\end{eqnarray}
From the theorem we thus know that there exists a $p^2$ such that\\[-5pt]
\begin{equation}
\label{pitay}
p^2\;=\;p^2_{\infty}+\sum_{k=1}^{\infty} \frac{G^k_{N}}{k!}  \left.\frac{d^{k}p^2}{dG_N^k}\right|_{G_N=0}\,, \quad p^2_{\infty}=\frac{(m_1^2+m_2^2-E^2)^2-4m_{1}^{2}m_{2}^{2}}{4E^2}\,,
\end{equation}
where the first term is nothing else than the solution to eq. (\ref{cons}) in the absence of interactions. The next term can be found using eqs. (\ref{der1}) and (\ref{der2}), giving\\[-5pt]
\begin{equation}
\label{p1}
\left.\frac{dp^2}{dG_N}\right|_{G_N=0} = -\left.\frac{\partial_{G_N}V}{\frac{1}{2 E \xi}+\partial_{p^2}V}\right|_{G_N=0}
= -2 E \xi \bigg[\frac{c_1}{r} \bigg]_{|_{p=p_{\infty}}}\,,
\end{equation}
\begin{equation}
\label{p2}
\left.\frac{d^2p^2}{d^2G_N}\right|_{G_N=0} = -2 E \xi \bigg[\frac{2c_2}{r^2}-\frac{4E\xi\, c_1 \partial_{p^2}c_1}{r^2}+\frac{c_1^2}{r^2}\bigg(\frac{3 \xi -1}{E \xi} \bigg) \bigg]_{|_{p=p_{\infty}}}\,,
\end{equation}
and so on for higher derivatives.\\[20pt]
Apparently, the structure of the $k$-derivative of $p^2$ as a function of $G_N$ seems to show no discernible structure, involving the potential and its derivatives. However, almost unbelievably, precisely the same relations also appear in the classical part of the position-space representation of the Lippmann-Schwinger equation that we have just examined above. There they relate the classical part of an $n$-loop amplitude to the potential and its derivatives. Indeed, by substituting eqs. (\ref{p1}) and (\ref{p2}) into eq. (\ref{pitay}), we see that the derivatives of $p^2$ satisfies a remarkable relation to the classical part of the position-space representation of loop amplitudes:
\vspace{2mm}\\[-5pt]
\begin{equation}
G_{N} \left.\frac{dp^2}{dG_N}\right|_{G_N=0}\;=\;-2 E \xi \bigg[\mathcal{\tilde{M}}^{cl.}_{\it tree}(p^2_{\infty},r) \bigg]\,,
\end{equation}
\begin{equation}
\left.\frac{G^2_{N}}{2}\frac{d^2p^2}{d^2G_N}\right|_{G_N=0}\;=\;-2 E \xi \bigg[\mathcal{\tilde{M}}^{cl.}_{\it 1-loop}(p^2_{\infty},r) \bigg]\,.
\end{equation}
By substituting these into eq. (\ref{pitay}), we observe that the implicit function we were searching for is precisely the classical part of the Fourier transform of the scattering amplitude,\\[-5pt]
\begin{equation}
\label{Zvi}
p^2\;=\;p^2_{\infty}-2 E \xi \bigg[\mathcal{\tilde{M}}^{cl.}_{\it tree}(p^2_{\infty},r)+\mathcal{\tilde{M}}^{cl.}_{\it 1-loop}(p^2_{\infty},r)\bigg]+\ldots\,.
\end{equation}
Indeed, the correspondence between solutions to the classical part of the Lippmann-Schwinger equation and the relativistic energy relation is not a coincidence and can be generalized to any loop order. The validity of eq. (\ref{Zvi}) is a consequence of Dini's theorem which maps the implicit function $p^2$ of the relativistic energy equation to the solution of the classical part of the Lippmann-Schwinger equation in position space.\\[5pt] 
The same relation\footnote{{As discussed in ref. \cite{Kalin}, the inclusion of radiative effects introduce a non-linear relation between $p^2$ and scattering amplitudes. These enter at 4PM order for a non-spinning binary system \cite{Berndue}. Our analysis is valid only in the conservative sector of the two body problem.}} \cite{Berndue,Kalin},\\[-10pt] 
\begin{equation}
\label{formula}
p^2=p^2_{\infty}-2 E\xi \mathcal{\tilde{M}}^{cl.}(p_{\infty},r) \,  ,
\end{equation}
can also be inferred by an intriguing alternative route suggested by Damour \cite{Damourdue} and recently generalized to all orders in by K\"alin and Porto in \cite{Kalin}. We rephrase it as follows.
Consider the energy equation in a fully relativistic system subjected to a post-Minkowskian potential. In the center of mass frame,\\[-10pt]
\begin{eqnarray}
\label{firstqua}
E &\;=\;& \sum_{i=1}^{2}\sqrt{p^2+m^2_i}+V(p,r)\,, \quad V(p,r)=\sum_{n=1}^{\infty}\frac{G_{N}^nc_{n}(p^2)}{r^n}\,, \label{first} 
\end{eqnarray}
\begin{equation}
\label{second}
p^2=p^2_{\infty}+\sum_{n=1}^{\infty}\frac{G_{N}^nf_{n}(E)}{r^n}\,,
\end{equation}
where eq. (\ref{second}) provides the perturbatively expanded solution to the energy condition and the $f_n$ coefficients that can be determined order by order in the coupling constant. A natural quantization of this \cite{Cristofoliuno} is the Salpeter Hamiltonian of relativistic particle states (\ref{opeH}),\\[-10pt]
\begin{equation}
\hat{H}\;=\;\sum_{i=1}^{2}\sqrt{\hat{p}^2+m^2_i}+\hat{V}\,,
\end{equation}
from which we infer the Lippmann-Schwinger equation discussed above. Given the nature of this Hamiltonian, it comes as no surprise that the associated Green function will have an intricate structure involving square roots as we have discussed in the previous section. Damour \cite{Damourdue} considers instead the second relation (\ref{second}) as a formally non-relativistic energy relation for a particle of mass 1/2 in
appropriate units. Because there is a map from eq. (\ref{first}) to eq. (\ref{second}) it should be equally meaningful to quantize
the Hamiltonian in $p^2$ of eq. (\ref{second}) as the original Salpeter Hamiltonian (\ref{first}). This means that we can use
a much simpler non-relativistic Hamiltonian to derive relations for the scattering amplitude. Its potential depends only on the radial distance $r$ as we are familiar with in ordinary non-relativistic quantum mechanics. We will thus
have all the powerful technology of non-relativistic quantum mechanics (and classical mechanics) at our disposal.\\[5pt]
The scattering amplitude will not be normalized as the original one, but this is of no immediate concern since physical observables should not 
depend on it as long as we rescale units appropriately. Damour's effective Hamiltonian operator is thus
\begin{equation}
\label{potef}
\mathcal{\hat{H}}\;=\;\hat{p}^2+V_{\it eff}(r) \ , \quad V_{\it eff}(r)\;\equiv-\;\sum_{n=1}^{\infty}\frac{G^n_Nf_n(E)}{r^n}\,,
\end{equation} 
which is a simple non-relativistic system with a potential given by Newtonian-like contributions of $r$-dependence only. 
For such a system, the classical part of the associated Lippmann-Schwinger equation is trivial in $D=4$. Indeed, all energy denominators in the Born subtractions will be just quadratic in the momenta and since the associated potential has no momentum-dependence, there is no expansion that could lead to classical terms. We thus find that the effective potential $V_{\it eff}(r)$ to all orders is proportional to the Fourier transform of the classical part of the corresponding amplitude evaluated at $p_{\infty}$, as before.\\[5pt]
We note that the $f_{n}(E)$ coefficients of eq. (\ref{potef}) are proportional to the classical part of the Fourier transform of the amplitude, $viz.$,
\begin{equation}
\label{fcoefficients}
\mathcal{\tilde{M}}^{cl.}(p,r) \equiv -\frac{1}{2 E\xi} \sum_{n=1}^{\infty}\frac{G_N^{n}\tilde{c}_{(n-1)-loop}(p)}{r^n} \quad \Rightarrow \quad f_{n}(E)=\tilde{c}_{(n-1)-loop}(p_{\infty}) \, .
\end{equation}
As we will see, these coefficients lead directly to the post-Minkowskian scattering angle in the center of mass frame. 
\section{The scattering angle to all orders}
The computation of the scattering angle for non-relativistic quantum mechanical Hamiltonians has a long history. Typically, 
interest has been mainly on finding approximate (semi-classical) solutions, first through the WKB-approximation, later by
considering the eikonal limit (see, {\it e.g.}, refs. \cite{Sugar,Paliov,Wallace}). These methods are powerful, but
they quickly get complicated and they were, of course, developed as approximate solutions to the full quantum mechanical
problem. \\[5pt]
Armed with the map of Hamiltonians from (\ref{first}) to (\ref{second}) we are in a completely different situation since
we can treat (\ref{second}) as a quantum mechanical Hamiltonian from which we only wish to extract the classical part. Not only
is the problem then WKB-exact, it is also WKB-trivial in the sense that we only wish to retain the leading $\hbar^0$-piece
of the wave function. This leading term $S$, as is well known, is a solution of the classical Hamilton-Jacobi equation. At
this stage we have therefore come full circle and we are back at analyzing the classical Hamiltonian (\ref{second}) with the
added knowledge that $f_{n}$ coefficients are simply identified with the Fourier transformed scattering amplitude evaluated at $p_{\infty}$
as seen from (\ref{fcoefficients}).\\[5pt]
Using this observation, we now provide an all-order expression for the post-Minkowskian scattering angle only in terms of the classical part of the amplitude in position space and the impact parameter $b$, both gauge invariant quantities.\\ [5pt]
As is well known that scattering angle is given from Hamilton-Jacobi theory by
\begin{equation}
\frac{\chi}{2}\;=\;- \int_{r_{m}}^{+\infty}dr~ \frac{\partial p_r}{\partial L}-\frac{\pi}{2}\,,
\end{equation}
where
\begin{equation}
\label{pir}
p_{r}\;=\;\sqrt{p^2_{\infty}-\frac{L^2}{r^2}-V_{\it eff}(r)} \ , \quad V_{\it eff}(r)\;=\;-\sum_{n=0}^{\infty}\frac{G_{N}^{n}f_{n}(E)}{r^n}\,,
\end{equation}
being $L$ the angular momentum of the system and $r_{m}$ the closest root to the origin of eq. (\ref{pir}) which satisfies
\begin{equation}
\label{rmincond}
1-\frac{b^2}{r^2_{m}}-\frac{V_{\it eff}(r_m)}{p^2_{\infty}}\;=\;0 \,, \quad b=\frac{L}{p_{\infty}}\,,
\end{equation}
where we have introduced the impact parameter $b$.\\
We find it convenient to rewrite the scattering angle as
\begin{equation}
\frac{\chi}{2}\;=\;b\int_{r_{m}}^{+\infty}\frac{dr}{r^2} \:\bigg(1-\frac{b^2}{r^2}-V_{\it eff}(r) \bigg)^{-\frac{1}{2}}-\frac{\pi}{2}
\;=\;b\int_{r_m}^{+\infty}\frac{dr}{r^2}\:\bigg(1-\frac{r^2_m}{r}-W(r) \bigg)^{-\frac{1}{2}}-\frac{\pi}{2}\,,
\end{equation}
where we have defined
\begin{equation}
\label{wdefined}
W(r)\;\equiv\;\frac{1}{p^2_{\infty}}\bigg[V_{\it eff}(r)-\frac{r^2_m}{r^2}V_{\it eff}(r_m)\bigg] \ , \quad W(r_m)\;=\;0\,.
\end{equation}
We next perform a change of variables to highlight the properties of $W(r)$ at $r_m$,
\begin{equation}
\label{change}
r^2 \;=\; u^2+r^2_{m} \quad \Rightarrow \quad \frac{\chi}{2}\;=\;b\int_{0}^{+\infty}\frac{du}{r^2}\bigg(1-\frac{r^2 W(r)}{u^2} \bigg)^{-\frac{1}{2}}-\frac{\pi}{2}\,.
\end{equation}
At this point we expand the square root of eq. (\ref{change}) using the generalized binomial theorem
\begin{equation}
\label{scuare}
(1+x)^{-\frac{1}{2}}\;=\;1+\sum_{n=0}^{\infty}\binom{-\frac{1}{2}}{n+1}x^{n+1}\,,
\end{equation}
where
\begin{equation}
\binom{-\frac{1}{2}}{n+1}\;=\;\frac{\Gamma(\frac{1}{2})}{\Gamma(n+2)\Gamma(-n-\frac{1}{2})}\;=\;\frac{(-1)^{n+1}(2n+1)!!}{2^{n+1}\Gamma(n+2)}\,.
\end{equation}
Using eq. (\ref{scuare}) the scattering angle becomes
\begin{equation}\begin{split}
\frac{\chi}{2}&\;=\;\frac{\pi}{2}\bigg(\frac{b}{r_m}-1\bigg)\;+\;b\sum_{n=0}^{\infty}(-1)^{n+1}\binom{-\frac{1}{2}}{n+1}\int_{0}^{+\infty}\frac{du}{u^{2(n+1)}}[W^{n+1}(r)r^{2n}]\\
&\;=\;\frac{\pi}{2}\bigg(\frac{b}{r_m}-1\bigg)\;+\;b\sum_{n=0}^{\infty}\frac{(2n+1)!!}{2^{n+1}(n+1)!}\int_{0}^{+\infty}\frac{du}{u^{2(n+1)}}[W^{n+1}(r)r^{2n}]\,.
\end{split}\end{equation}
We now use the following properties which holds for $\mathcal{C}^{\infty}$ functions from $\mathbb{R}$ to $\mathbb{R}$ that vanish at infinity and at the origin:
\begin{equation}
\label{parts}
\int_{0}^{+\infty}\frac{du}{u^{2(n+1)}}f(u)\;=\;\frac{1}{(2n+1)!!}\int_{0}^{+\infty}du \bigg(\frac{1}{u}\frac{d}{du}\bigg)^{n+1}f(u) \,.
\end{equation}
Using eq. (\ref{parts}) we obtain
\begin{equation}\begin{split}
\frac{\chi}{2}&\;=\;\frac{\pi}{2}\bigg(\frac{b}{r_m}-1\bigg)+b\sum_{n=0}^{\infty}\frac{1}{2^{n+1}(n+1)!}\int_{0}^{+\infty} du \bigg(\frac{1}{u}\frac{d}{du}\bigg)^{n+1}\big[W^{n+1}(r)r^{2n}\big]\\
&\;=\;\frac{\pi}{2}\bigg(\frac{b}{r_m}-1\bigg)+b\sum_{n=0}^{\infty}\frac{1}{(n+1)!}\int_{0}^{+\infty} du \bigg(\frac{d}{du^2}\bigg)^{n+1}\big[W^{n+1}(r)r^{2n}\big]\,.
\end{split}\end{equation}
In order to introduce a systematic expansion we write this as\\[-0.2cm]
\begin{equation}
\frac{\chi}{2}\;=\;\frac{\pi}{2}\bigg(\frac{b}{r_m}-1\bigg)\;+\;b\sum_{n=0}^{\infty}\Delta_{n}(r_m)\,,
\end{equation}\vskip-0.2cm\noindent
\begin{equation}
\label{deltaz}
\Delta_{n}(r_m) \;\equiv\; \frac{1}{(n+1)!}\int_{0}^{+\infty}du \bigg(\frac{d}{du^2}\bigg)^{n+1}\big[W^{n+1}(r)r^{2n}\big] \ , \quad r\;=\;\sqrt{u^2+r^2_m}\,.
\end{equation}
Focusing on eq. (\ref{deltaz}), we now expand
\begin{equation}
\Delta_{n}(r_m) \;=\; \frac{1}{p^{2n+2}_{\infty}}\frac{1}{(n+1)!}\sum_{k=0}^{n+1}\binom{n+1}{k}\!\int_{0}^{+\infty}\!\!du \bigg(\frac{d}{du^2}\bigg)^{n+1}\!\!\big[V_{\it eff}^{n+1-k}(r)r^{2n}\big]\bigg[-\frac{r^2_{m}V_{\it eff}(r_{m})}{r^2}\bigg]^{k}\,.
\end{equation}
Rewriting in terms of $b$ and $r_m$, and using eq. (\ref{rmincond}), this leads to
\begin{eqnarray}\displaystyle
\Delta_{n}(r_m)&&\;=\;\sum_{k=0}^{n+1}\frac{(b^2-r^2_m)^k}{k!}\int_{0}^{+\infty}du \bigg(\frac{d}{du^2}\bigg)^{n+1}\frac{V_{\it eff}^{n-k+1}(r)\,r^{2(n-k)}}{(n-k+1)!\,p^{2(n-k+1)}_{\infty}}
 \label{de}\\
&&\;=\;\sum_{k=0}^{n+1}\frac{(b^2-r^2_m)^k}{k!}\bigg(\frac{d}{dr_{m}^2}\bigg)^{k}\int_{0}^{+\infty}du \bigg(\frac{d}{du^2}\bigg)^{n-k+1}\frac{V_{\it eff}^{n-k+1}(r)\,r^{2(n-k)}}{(n-k+1)!\,p^{2(n-k+1)}_{\infty}}\,,\nonumber
\end{eqnarray}
where we have used the fact that derivatives on $r_m^2$ and $u^2$ can be interchanged for a function of the radial distance $r=\sqrt{u^2+r^2_m}$, so as to put these outside the integration. This simple trick, allows us to recognize in eq. (\ref{de}) the following function
\begin{equation}
\chi_{m}(r_m) \;\equiv\; \frac{1}{p^{2m+2}_{\infty}}\int_{0}^{+\infty}du \bigg(\frac{d}{du^2}\bigg)^{m+1}\frac{V_{\it eff}^{m+1}(r)\,r^{2m}}{(m+1)!}\,,
\end{equation}
using which we can rewrite eq. (\ref{de}) as
\begin{equation}
\Delta_{n}(r_m)\;=\;\sum_{k=0}^{n+1}\tilde{\Delta}_{n,k}(r_m) \ , \quad \tilde{\Delta}_{n,k}(r_m)\;\equiv\; \frac{(b^2-r^2_m)^k}{k!}\bigg(\frac{d}{dr_{m}^2}\bigg)^{k}\chi_{n-k}(r_m)\,.
\end{equation}
To summarize what we have obtained so far,
\begin{eqnarray}
\frac{\chi}{2} &\;=\;& \frac{\pi}{2}\bigg(\frac{b}{r_m}-1\bigg)\;+\;b\sum_{n=0}^{\infty}\sum_{k=0}^{n+1}\tilde{\Delta}_{n,k}(r_m) \cr
&\;=\;& \frac{\pi}{2}\bigg(\frac{b}{r_m}-1\bigg)\;+\;b\sum_{n=0}^{\infty}\sum_{k=0}^{n}\tilde{\Delta}_{n,k}(r_m)\;+\;b\sum_{n=0}^{\infty}\tilde{\Delta}_{n,n+1}(r_m)\,. \label{sumsca}
\end{eqnarray}
The last sum can be rewritten in a remarkably simple way
\begin{eqnarray}
b \sum_{n=0}^{\infty}\tilde{\Delta}_{n,n+1}(r_m) &=& b\, \sum_{n=0}^{\infty}\frac{(b^2-r^2_m)^{n+1}}{(n+1)!}\bigg(\frac{d}{dr_{m}^2}\bigg)^{n+1}\!\!\!\!\chi_{-1}(r_m)\!\!\!\! \\
&=& b\,\sum_{n=0}^{\infty}\!\frac{(b^2-r^2_m)^{n}}{n!}\bigg(\frac{d}{dr_{m}^2}\bigg)^{n}\!\!\chi_{-1}(r_m)-b \chi_{-1}(r_m) 
= b\,\big[\chi_{-1}(b)-\chi_{-1}(r_m)\big]\,,\nonumber
\end{eqnarray}
or simply
\begin{equation}
b \sum_{n=0}^{\infty}\tilde{\Delta}_{n,n+1}(r_m)\;=\;\frac{\pi}{2}\bigg(1-\frac{b}{r_m}\bigg)\,, \label{Taylormagic}
\end{equation}
where we have  recognized the Taylor series of $\chi_{-1}(r_m)$ around $b$. This is equal and opposite to the first contribution of eq. (\ref{sumsca}), a cancellation which lead to the following expression for the scattering angle
\begin{equation}
\frac{\chi}{2}\;=\;b\sum_{n=0}^{\infty}\sum_{k=0}^{n}\tilde{\Delta}_{n,k}(r_m)
\label{taylorma2}
\;=\;b\sum_{n=0}^{\infty}\sum_{k=0}^{n}\frac{(b^2-r^2_m)^k}{k!}\bigg(\frac{d}{dr_{m}^2}\bigg)^{k}\chi_{n-k}(r_m)
\;=\;b\sum_{k=0}^{\infty}\chi_{k}(b) ~.
\end{equation}
In the last equality we have used the fact that eq. (\ref{taylorma2}) is the sum over $n$ of the Taylor series of $\chi_{n}(r_m)$ around $b$. Thus, the main result of this section can be summarized in the following way, which states that the scattering angle can always be expressed in terms of finite integrals without any reference to $r_{m}$
\begin{equation}
\chi\;=\;\sum_{k=1}^{\infty}\tilde{\chi}_{k}(b) \ ,\quad \tilde{\chi}_{k}(b)\;\equiv\;  \frac{2b}{k!}\int_{0}^{+\infty}du \bigg(\frac{d}{du^2}\bigg)^{k}\frac{V_{\it eff}^{k}(r)\,r^{2(k-1)}}{p^{2k}_{\infty}}\,. \label{finalchi}
\end{equation}
Since $V_{\it eff}$ is related to the classical part of the Fourier transform of scattering amplitudes from eq. (\ref{fcoefficients}), this concludes the derivation of the scattering angle solely in terms of gauge invariant quantities.
The manifest independence of the intermediate parameter $r_m$ (the distance of nearest approach) in our expression for the
scattering angle is important. Since $r_m$ in general is determined by a solvable condition relating it to other scattering
information it should disappear entirely from the result, as we have shown explicitly. In our approach there is no subtlety
involved in the way it drops out of the relation for the scattering angle and there is no need to regularize intermediate
expressions on account of it. Independence of $r_m$ is a particularly acute problem in general relativity where this quantity is not even gauge invariant and such it has to disappear from the expression for the gauge invariant scattering angle.\\ [5pt]
Let us finally explore the simplicity of our expression for the scattering angle as opposed to previous methods. As described above, we can express the fully relativistic scattering angle in terms of an effective position-space potential which for the case of general relativity is given by
\begin{equation}
V_{\it eff}(r)\;=\;-\sum_{n=1}^{\infty}\frac{G_N^{n}f_{n}(E)}{r^n}\,.
\end{equation}
This is related to the classical part of the scattering amplitude to any loop order as shown. Let us first focus on the angle up to 3PM order in four dimensions, later generalizing it to all dimensions. We thus consider
\vspace{2mm}
\begin{eqnarray}
\label{3pm}
\chi^{3PM}(b)\;&=&\;\tilde{\chi}_{1}(b)+\tilde{\chi}_{2}(b)+\tilde{\chi}_{3}(b)\,,\\[5pt]\nonumber
\tilde{\chi}_{1}(b)\;&=&\;\frac{2b}{p^2_{\infty}}\int_{0}^{+\infty}du \:\frac{d}{db^2}V_{\it eff}(r)\,,\\
\label{cai2}
\tilde{\chi}_{2}(b)\;&=&\;\frac{b}{p^{4}_{\infty}}\int_{0}^{+\infty}du \:\bigg(\frac{d}{db^2}\bigg)^{2} r^2\big[V_{\it eff}(r)\big]^{2}\,,\\
\label{cai3}
\tilde{\chi}_{3}(b)\;&=&\;\frac{b}{3p^{6}_{\infty}}\int_{0}^{+\infty}du \:\bigg(\frac{d}{db^2}\bigg)^{3}r^4\big[V_{\it eff}(r)\big]^{3}\,.
\end{eqnarray}
We start with the first contribution from eq. (\ref{3pm}), 
\begin{equation}
\label{bom}
\tilde{\chi}_{1}(b)\;=\;\frac{b}{p^2_{\infty}} \int_{0}^{+\infty} du \: \frac{\partial_{r}V_{\it eff}(\sqrt{u^2+b^2})}{\sqrt{u^2+b^2}} \,.
\end{equation}
This we recognize as a classic textbook formula, usually presented for the bending angle around static massive sources in the non-relativistic approximation (see, {\it e.g.}, ref.~\cite{Bohm}). Although it is surely of older origin, we will denote it Bohm's formula.
The power of our derivation is that this formula describes the motion of fully relativistic particles, with no restriction on masses or range of velocities on account of the exact map. We can also provide a closed formula for this contribution given by a generic effective potential
\begin{equation}
\tilde{\chi}_{1}(b)\;=\;\frac{b}{p^2_{\infty}} \sum_{n=1}^{\infty}nG^{n}_{N}f_{n}(E)\int_{0}^{+\infty} du \frac{1}{(u^2+b^2)^{\frac{n}{2}+1}}\,.
\end{equation}
As can be seen, all terms depend on the integral
\begin{equation}
\label{integr}
\int_{0}^{+\infty} du \frac{1}{(u^2+b^2)^{\frac{n}{2}+1}}\;=\;\frac{1}{b^{n+1}}\frac{\sqrt{\pi}}{n}\frac{\Gamma(\frac{n+1}{2})}{\Gamma(\frac{n}{2})}\,, \quad \forall n \in \mathbb{N}\,,
\end{equation}
and thus
\begin{equation}
\tilde{\chi}_{1}(b)\;=\;\frac{\sqrt{\pi}}{p^2_{\infty}} \sum_{n=1}^{\infty}\frac{G^{n}_{N}f_{n}(E)}{b^{n}}\frac{\Gamma(\frac{n+1}{2})}{\Gamma(\frac{n}{2})}\,.
\end{equation}
To 3PM order, the other needed contributions are given by
\begin{equation}
\label{lin}
\tilde{\chi}_{1}(b)\;=\;\frac{\sqrt{\pi}}{p^2_{\infty}} \sum_{n=1}^{3}\frac{G^{n}_{N}f_{n}(E)}{b^{n}}\frac{\Gamma(\frac{n+1}{2})}{\Gamma(\frac{n}{2})}=\frac{G_Nf_1}{Lp_{\infty}}+\frac{G_{N}^2 f_2 \pi}{2 L^2}+\frac{G_{N}^3f_32p_{\infty}}{L^3}\,,
\end{equation}
which reproduces the linear terms in $f_{n}$ up to 3PM known in literature. However, to the same order there are also additional contributions which can be regarded as corrections to Bohm's formula beyond leading order as given by eqs. (\ref{cai2})-(\ref{cai3})
\begin{eqnarray}
\tilde{\chi}_{2}(b)\;&=&\;\frac{b}{p^4_{\infty}}\int_{0}^{+\infty}du \: \bigg(\frac{d}{db^2} \bigg)^2 r^2\bigg[\frac{2G^3_{N}f_1 f_2}{r^3}+\frac{G^2_{N}f_1^2}{r^2} \bigg] 
= \frac{G^3_{N}f_1 f_2 }{L^3 p_{\infty}}\,.
\end{eqnarray}
Here is an important observation: The contribution to $G_N^2$ vanishes in four dimensions.
This means that Bohm's formula in eq. (\ref{bom}) is valid, beyond what we could expect, also at 2PM order, a fact which has been previously noticed and from which now we provide a clear understanding. In fact, Bohm's non-relativistic formula holds at 2PM order even if one 
naively substitutes a static non-relativistic potential for the bending of light \cite{Bjerrum-Bohruno,Bjerrum-Bohrtre}, and we now
understand why. Furthermore, this formula agrees with the explicit calculations of the eikonal limit of gravity up to 2PM order with
arbitrary masses \cite{'tHooft,Kabat,Sterman,Bjerrum-Bohruno,Paolouno,Paolodue}. We now also understand why the eikonal exponentiation
of classical gravity works out so simply at 2PM order in four dimensions: it is the vanishing of the $f_1^2$-term for the angle (and
the fact that in the eikonal limit the scattering angle enters in terms of the odd function $\sin(\chi)$).\\[5pt]
This brings us to another important point. We see from this analysis that the eikonal exponentiation is bound to work for
classical gravity to all orders and in any number of dimensions. Not only that, its precise form is already dictated by
the formula we provide. In this sense, there would superficially seem to be no need to pursue the computation of the 
eikonal limit beyond 2PM order. However, given that the actual evaluation
of the coefficients $f_i$ require explicit full amplitude calculations it could still be of interest to pursue the
eikonal limit to the given order, as an independent check. \\[5pt]
Finally, we need to evaluate the remaining term
\begin{equation}
\label{cub}
\tilde{\chi}_{3}(b)\;=\;-\frac{bG^{3}_{N}f^{3}_{1}}{3p^{6}_{\infty}}\int_{0}^{+\infty}du\: \bigg(\frac{d}{db^2} \bigg)^{3}r=-\frac{G^3_{N}f^3_{1}}{12L^3p^{3}_{\infty}}\,.
\end{equation}
Summing these contributions, we obtain the desired scattering angle at 3PM order
\begin{eqnarray}
\chi^{3PM}\;&=&\;\frac{G_Nf_1}{Lp_{\infty}}+\frac{G_N^2 f_2 \pi}{2 L^2}+\frac{G_N^3f_32p_{\infty}}{L^3}+\frac{G^3_{N}f_1 f_2 }{L^3 p_{\infty}}-\frac{G^3_{N}f^3_{1}}{12L^3p^{3}_{\infty}},\\
\;&=&\;\frac{G_N\,\tilde{c}_{\it tree}}{Lp_{\infty}}+\frac{G_N^2 \,\tilde{c}_{\it 1-loop} \pi}{2 L^2}+\frac{G_N^3\,\tilde{c}_{\it 2-loop}\,2p_{\infty}}{L^3}+\frac{G^3_{N}\,\tilde{c}_{\it tree} \,\tilde{c}_{\it 1-loop}}{L^3 p_{\infty}}-\frac{G^3_{N}\tilde{c}^3_{\it tree}}{12L^3p^{3}_{\infty}}\,.\nonumber
\end{eqnarray}
As seen, the computation is quite straightforward, involving only elementary integrals and derivatives.  Higher PM contributions can be calculated easily to any desired order as demonstrated in table \ref{tab:mytable}.
It is clear that there are interesting patterns in these expressions and it is elementary to express several of the
combinations in simple closed form, valid to all orders.\\[5pt]
It is perhaps more interesting to note that certain combinations are {\em missing}. We illustrated this above by pointing
out how the $f_1^2$-contribution vanishes. 
Equipped with the map between effective potential and coefficient of the amplitude, we can now understand this 
result in all generality.\footnote{{\footnotesize We thank R. Porto for pointing out that the condition for vanishing contributions given in the first version of this paper was sufficient but not necessary.}}\\[5pt]
In order to analyse the general conditions for such vanishing contribution to the scattering angle 
we start by reconsidering the previous expression for the post-Minkowskian scattering angle assuming a $m$ post-Minkowskian potential
\begin{equation}
\label{veff}
\chi=\sum_{n=1}^{\infty}\tilde{\chi}_{n}(b)\,, \quad \tilde{\chi}_{n}(b)=\frac{2b}{n!\, p^{2n}_{\infty}}\int_{0}^{+\infty}du \: \bigg(\frac{d}{db^2}\bigg)^{n}r^{2n-2}V_{\it eff}^{n}(r)\,,
\end{equation}
\begin{equation}
V_{\it eff}(r)=-\sum_{k=1}^{m}\frac{G_N^{k}f_k}{r^k}\,.
\end{equation}
\begin{table}[hhh]\small\label{Table1}
\begin{equation}
\begin{array}{|c|l|l}\hline\displaystyle
{\rm PM }& {\rm \chi^{\rm PM}}/ \big( {G_N\over p_{\infty} L} 
\big)^{\rm PM }\\[2pt] \hline
1 &f_1\\[2pt] \hline
2&\frac{1}{2} \pi  p_{\infty }^2  f_2\\ \hline
3&2 f_3 p_{\infty }^4+f_1 f_2 p_{\infty }^2-\frac{f_1^3}{12}\\ \hline
4&\frac{3}{8} \pi  p_{\infty }^4 \big(2 f_4 p_{\infty }^2+f_2^2+2 f_1
    f_3\big)\\ \hline
5&\frac{8}{3} f_5 p_{\infty }^8+4 \big(f_2 f_3+f_1 f_4\big) p_{\infty 
}^6+f_1
    \big(f_2^2+f_1 f_3\big) p_{\infty }^4-\frac{1}{6} f_1^3 f_2 p_{\infty
    }^2+\frac{f_1^5}{80}\\ \hline
6&\frac{5}{16} \pi  p_{\infty }^6 \big(3 f_6 p_{\infty }^4+3 
\big(f_3^2+2 f_2 f_4+2 f_1
    f_5\big) p_{\infty }^2+f_2^3+6 f_1 f_2 f_3+3 f_1^2 f_4\big)\\ \hline
7&\frac{16}{5} f_7 p_{\infty }^{12}+8 \big(f_3 f_4+f_2 f_5+f_1 f_6\big) 
p_{\infty
    }^{10}+6 \big(f_3 f_2^2+2 f_1 f_4 f_2+f_1 \big(f_3^2+f_1 f_5\big)\big)
    p_{\infty }^8\\&+f_1 \big(f_2^3+3 f_1 f_3 f_2+f_1^2 f_4\big) p_{\infty
    }^6-\frac{1}{8} f_1^3 \big(2 f_2^2+f_1 f_3\big) p_{\infty 
}^4+\frac{3}{80} f_1^5
    f_2 p_{\infty }^2-\frac{f_1^7}{448}\\ \hline
8&\frac{35}{128} \pi  p_{\infty }^8 \big(4 f_8 p_{\infty }^6+6 
\big(f_4^2+2 \big(f_3
    f_5+f_2 f_6+f_1 f_7\big)\big) p_{\infty }^4+12 \big(f_4 
f_2^2+\big(f_3^2+2 f_1
    f_5\big) f_2\\ &+f_1 \big(2 f_3 f_4+f_1 f_6\big)\big) p_{\infty 
}^2+f_2^4+6 f_1^2
    f_3^2+12 f_1 f_2^2 f_3+12 f_1^2 f_2 f_4+4 f_1^3 f_5\big)\\ \hline
    9&\frac{128}{35} f_9 p_{\infty }^{16}+\frac{64}{5} \big(f_4 f_5+f_3 
f_6+f_2 f_7+f_1
    f_8\big) p_{\infty }^{14}+\frac{16}{3} \big(f_3^3+6 \big(f_2 f_4+f_1 
f_5\big)
    f_3+3 f_2^2 f_5\\ &+3 f_1 \big(f_4^2+2 f_2 f_6+f_1 f_7\big)\big) 
p_{\infty }^{12}+8
    \big(f_3 f_2^3+3 f_1 f_4 f_2^2+3 f_1 \big(f_3^2+f_1 f_5\big) f_2\\ 
&+f_1^2 \big(3
    f_3 f_4+f_1 f_6\big)\big) p_{\infty }^{10}+f_1 \big(f_2^4+6 f_1 f_3 
f_2^2+4
    f_1^2 f_4 f_2+f_1^2 \big(2 f_3^2+f_1 f_5\big)\big) p_{\infty }^8\\ 
&-\frac{1}{30}
    f_1^3 \big(10 f_2^3+15 f_1 f_3 f_2+3 f_1^2 f_4\big) p_{\infty 
}^6+\frac{1}{40}
    f_1^5 \big(3 f_2^2+f_1 f_3\big) p_{\infty }^4-\frac{1}{112} f_1^7 
f_2 p_{\infty
    }^2+\frac{f_1^9}{2304}\\ \hline
    10&\frac{63}{256} \pi  p_{\infty }^{10} \big(5 f_{10} p_{\infty 
}^8+10 \big(f_5^2+2
    \big(f_4 f_6+f_3 f_7+f_2 f_8+f_1 f_9\big)\big) p_{\infty }^6\\ &+30 
\big(f_6
    f_2^2+\big(f_4^2+2 f_1 f_7\big) f_2+f_3^2 f_4+2 f_3 \big(f_2 f_5+f_1
    f_6\big)+f_1 \big(2 f_4 f_5+f_1 f_8\big)\big) p_{\infty }^4\\ &+10 
\big(2 f_4
    f_2^3+3 \big(f_3^2+2 f_1 f_5\big) f_2^2+6 f_1 \big(2 f_3 f_4+f_1 
f_6\big)
    f_2+f_1 \big(2 f_3^3+6 f_1 f_5 f_3\\&+f_1 \big(3 f_4^2+2 f_1 
f_7\big)\big)\big)
    p_{\infty }^2+f_2^5+30 f_1^2 f_2 f_3^2+20 f_1 f_2^3 f_3+30 f_1^2 
f_2^2 f_4+20 f_1^3
    f_3 f_4\\&+20 f_1^3 f_2 f_5+5 f_1^4 f_6\big)\\ \hline
    11&\frac{256}{63} f_{11} p_{\infty }^{20}+\frac{128}{7} \big(f_5 
f_6+f_4 f_7+f_3 f_8+f_2
    f_9+f_1 f_{10}\big) p_{\infty }^{18}\\&+32 \big(f_7 f_2^2+2 \big(f_4 
f_5+f_1
    f_8\big) f_2+f_3^2 f_5+f_3 \big(f_4^2+2 f_2 f_6+2 f_1 
f_7\big)\\&+f_1 \big(f_5^2+2
    f_4 f_6+f_1 f_9\big)\big) p_{\infty }^{16}+\frac{80}{3} \big(f_8 f_1^3+3
    \big(f_4 \big(f_3^2+f_1 f_5\big)+f_1 f_3 f_6\big) f_1+f_2^3 f_5\\&+3 
f_2^2
    \big(f_3 f_4+f_1 f_6\big)+f_2 \big(f_3^3+6 f_1 f_5 f_3+3 f_1 
\big(f_4^2+f_1
    f_7\big)\big)\big) p_{\infty }^{14}+10 \big(f_3 f_2^4+4 f_1 f_4 
f_2^3\\&+6 f_1
    \big(f_3^2+f_1 f_5\big) f_2^2+4 f_1^2 \big(3 f_3 f_4+f_1 f_6\big) 
f_2+f_1^2
    \big(2 f_3^3+2 f_1 \big(f_4^2+2 f_3 f_5\big)+f_1^2 f_7\big)\big) 
p_{\infty
    }^{12}+\\&f_1 \big(f_2^5+10 f_1 f_3 f_2^3+10 f_1^2 f_4 f_2^2+5 f_1^2 
\big(2 f_3^2+f_1
    f_5\big) f_2+f_1^3 \big(5 f_3 f_4+f_1 f_6\big)\big) p_{\infty
    }^{10}\\&-\frac{1}{12} f_1^3 \big(5 f_2^4+15 f_1 f_3 f_2^2+6 f_1^2 
f_4 f_2+f_1^2
    \big(3 f_3^2+f_1 f_5\big)\big) p_{\infty }^8\\&+\frac{1}{56} f_1^5 
\big(7 f_2^3+7
    f_1 f_3 f_2+f_1^2 f_4\big) p_{\infty }^6-\frac{5}{896} f_1^7 \big(4 
f_2^2+f_1
    f_3\big) p_{\infty }^4+\frac{5 f_1^9 f_2 p_{\infty
    }^2}{2304}-\frac{f_1^{11}}{11264}\\ \hline
       12&\frac{231}{1024} \pi  p_{\infty }^{12} \big(6 f_{12} p_{\infty 
}^{10}+15 \big(f_6^2+2
    \big(f_5 f_7+f_4 f_8+f_3 f_9+f_2 f_{10}+f_1 f_{11}\big)\big) 
p_{\infty }^8\\&+20
    \big(f_4^3+6 \big(f_3 f_5+f_2 f_6+f_1 f_7\big) f_4+3 \big(f_8
    f_2^2+\big(f_5^2+2 f_3 f_7+2 f_1 f_9\big) f_2\\&+f_3^2 f_6+f_1 
\big(2 f_5 f_6+2 f_3
    f_8+f_1 f_{10}\big)\big)\big) p_{\infty }^6+15 \big(f_3^4+12 
\big(f_2 f_4+f_1
    f_5\big) f_3^2\\&+12 \big(f_5 f_2^2+f_1 \big(f_4^2+2 f_2 f_6+f_1 
f_7\big)\big)
    f_3+2 \big(2 f_6 f_2^3+3 \big(f_4^2+2 f_1 f_7\big) f_2^2\\&+6 f_1 
\big(2 f_4
    f_5+f_1 f_8\big) f_2+f_1^2 \big(3 f_5^2+6 f_4 f_6+2 f_1 
f_9\big)\big)\big)
    p_{\infty }^4\\&+30 \big(f_4 f_2^4+2 \big(f_3^2+2 f_1 f_5\big) 
f_2^3+6 f_1 \big(2
    f_3 f_4+f_1 f_6\big) f_2^2+2 f_1 \big(2 f_3^3+6 f_1 f_5 f_3\\&+f_1 
\big(3 f_4^2+2
    f_1 f_7\big)\big) f_2+f_1^2 \big(6 f_4 f_3^2+4 f_1 f_6 f_3+f_1 
\big(4 f_4
    f_5+f_1 f_8\big)\big)\big) p_{\infty }^2\\&+f_2^6+20 f_1^3 f_3^3+90 
f_1^2 f_2^2
    f_3^2+15 f_1^4 f_4^2+30 f_1 f_2^4 f_3\\&+60 f_1^2 f_2^3 f_4+120 
f_1^3 f_2 f_3 f_4+60
    f_1^3 f_2^2 f_5+30 f_1^4 f_3 f_5+30 f_1^4 f_2 f_6+6 f_1^5 f_7\big)\\ 
\hline     \end{array}\nonumber
\end{equation}
\caption{ PM corrections to 12th order in $G_N$.}
\label{tab:mytable}
\end{table}

We expand the $n$-power of the potential by using the multinomial theorem
\begin{equation}
V^{n}_{\it eff}(r)=(-1)^n\!\!\!\!\!\sum_{n_{1}+n_{2}+\ldots+n_{m}=n}\!\!\!\binom{n}{n_{1},n_{2},\ldots, n_{m}}\frac{G_{N}^{\beta_{m}}f_{1}^{n_1}f_{2}^{n_2}\cdots f_{m}^{n_{m}}}{r^{\beta_{m}}}\,,
\end{equation}
where $\beta_{m} \equiv n_{1}+2 n_{2}+3n_{3}+...m n_{m}$ and $\binom{n}{n_{1},n_{2},\ldots, n_{m}}\equiv\frac{n!}{n_{1}!n_{2}!...n_{m}!}$.
If we now evaluate eq.(\ref{veff}) using this we have
\begin{eqnarray}
\label{finalscattering}
\tilde{\chi}_{n}(b)\!&\!=\!\!&\!\frac{2\sqrt{\pi}}{n! p^{2n}_{\infty}}\sum_{n_{1}+n_{2}+\ldots+n_{m}=n}\!\!\!\!\!\!\!\!\frac{f_{1}\nonumber^{n_{1}}\!f_{2}^{n_2}\!\cdots f_{m}^{n_{m}}}{b^{\beta_{m}}}\frac{G_{N}^{\beta_{m}}}{\beta_{m}}\!\, \binom{n}{n_{1},n_{2},\ldots, n_{m}}\frac{\Gamma(\frac{\beta_{m}\!+\!1}{2})}{\Gamma(\frac{\beta_{m}}{2})}\! \prod_{\alpha=0}^{n-1}(1\!-\!n\!+\!\frac{\beta_{m}}{2}\!+\!\alpha)\\&=&\!\frac{\sqrt{\pi}}{p^{2n}_{\infty}}\sum_{n_{1}+n_{2}+\ldots+n_{m}=n}\bigg(\frac{G_{N}}{b}\bigg)^{\beta_{m}} \bigg(\prod_{l=1}^{m} \frac{f_{l}^{n_{l}}}{n_{l}!}\bigg) \frac{\Gamma(\frac{\beta_{m}\!+\!1}{2})}{\Gamma(\frac{\beta_{m}}{2}\!+\!1\!-\!n)}\,.
\end{eqnarray} 

Null contributions in eq.(\ref{finalscattering}) appears for%
\\[-0.4cm]
\begin{equation}
\begin{cases}
2n-2-n_1-2n_2-...-mn_m=0\,,\\
n_1+n_2+...n_m=n \,,  \quad  \forall n \wedge n_{j=1,2..m} \in \mathbb{N}\,,
\end{cases}
\end{equation}
\vspace{3mm}
\newline
as well as
\begin{equation}
\begin{cases}
2n-2-n_1-2n_2-...-mn_m-1=0\,,\\
n_1+n_2+...n_m=n\,, \quad  \forall n \wedge n_{j=1,2..m} \in \mathbb{N}\,,
\end{cases}
\end{equation}
and so on. All these can be expressed in a compact form as follows
\begin{equation}
\begin{cases}
2n-2-n_1-2n_2-...-mn_m-\alpha=0\,,\\
n_1+n_2+...n_m=n\,, \quad  \forall n,\alpha \wedge n_{j=1,2..m} \in \mathbb{N}\;\; :\;\; 0 \leq \alpha \leq n-1\,.
\end{cases}
\end{equation}
This system of equations describes the intersection of two affine hyperplanes in $m$ dimensions, the solutions to which are positive integer points on a parametric $m-2$ affine hyperplane with parameters $n$ and $\alpha$. Thus,
given a $m$-dimensional post-Minkowskian potential, the vanishing coefficients to the scattering angle are
in one-to-one correspondence with the positive integer zeros of the intersection of two affine hyperplanes in $m$ dimensions. As an example, let us evaluate the vanishing contributions to the scattering angle arising from a 3PM potential. The system to be solved is
\begin{equation}
\begin{cases}
n_1+2n_2+3n_3=2n-2-\alpha\,, \\
n_1+n_2+n_3=n\,, \quad  \forall n,\alpha \wedge n_{j=1,2,3} \in \mathbb{N} \;\; : \;\;  0 \leq \alpha \leq n-1\,,
\end{cases}
\end{equation}
and the solution is given by
\begin{equation}
\begin{cases}
n_1=n_1 \\
n_2=2-2n_1+n+\alpha\,, \\
n_3=n_1-2-\alpha\,,  \quad  \forall n,\alpha, n_{1} \in \mathbb{N}\;\; : \;\; 0 \leq \alpha \leq n-1\,.
\end{cases}
\end{equation}
We remind the reader that the parameter $n$ labels the $\tilde{\chi}_n$ contribution to the scattering angle. 
For $n=1$ there are no positive integer solution on this hyperplane, while for $n=2$ we find that there is only one solution given by $n_1=2,n_2=n_3=0$, which is nothing else than the vanishing of the $f_{1}^{n_1}=f^{2}_{1}$ term. This procedure is straightforward, it can be easily generalized to any order, and shows that there is an infinite number of such vanishing contributions.

\section{Conclusion}
We have unravelled an unexpected equivalence between classical solutions to Lippmann-Schwinger equations and solutions to the
relativistic energy relation of two-body dynamics. The equivalence ensures that a physical observable such as the scattering
angle can be determined directly from the classical part of the amplitude without recourse to the relativistic potential. In detail,
we have found that the implicit function theorem applied to the relativistic energy relation is in one-to-one correspondence
with the classical part of the solutions to the Lippmann-Schwinger equation of the quantum mechanical scattering problem. The link is
a relation between the classical part of the scattering amplitude and the potential (and derivatives thereof). Amazingly, this
relation removes all Born subtractions from the problem leaving us with only the classical part of the amplitude when we
evaluate the scattering angle.\\[5pt]
Using Damour's map to a non-relativistic theory for a particle of mass equal to 1/2, we have derived an
explicit formula for the Post-Minkowksian scattering angle to any order in the coupling constants of the potential. This formula
is universal and applicable to any classical potential. A distinct advantage of our formula is that it does not require
knowledge of the classical turning point $r_m$, nor does it require regularization with respect to that quantity.
When we apply our formula to the problem of Post-Minkowskian general relativity
we recover, effortlessly, the perturbative expansions quoted in the literature. We have illustrated the simplicity of our expression
for the scattering angle by listing the expression of the scattering angle up to 12PM order.\\[5pt]
There are patterns in these expressions for the scattering angle and we have explained why there are certain
``vanishing theorems'' for particular combinations of terms. 
The first missing one is the $f_1^2$-piece of the one-loop scattering angle, which explains the simplicity of the eikonal limit at one-loop order. We have also found the general condition for the vanishing of such contributions to any order.\\[5pt]
\subsection*{Acknowledgements}
We gratefully acknowledge numerous discussions with Ludovic Plant\'e and Pierre Vanhove on the derivation of the bending
angle from the eikonal limit. We would also like to thank Kays Haddad for conversations and for useful comments on the manuscript. This project has received funding from the European Union's Horizon 2020 research and innovation programme under the Marie 
Sklodowska-Curie grant agreement No. 764850 ``SAGEX'' and has also been supported in part by the Danish National Research Foundation (DNRF91). 

\begin{appendix}\section{Classical contributions from the Lippman-Schwinger equation}
In this Appendix we elaborate on the point made in the main text regarding the computational topology of the classical contributions from the 
Lippman-Schwinger equation. As we have seen in eq.(\ref{schw}), the n-th term of the Lippmann-Schwinger equation is given by
\begin{equation}
S_{n}(p,p')=\int_{k_1,k_2,\ldots,k_n}\frac{\tilde{V}(p,k_1)\tilde{V}(k_1,k_2)\cdots\tilde{V}(k_n,p')}{(E_p-E_{k_1})\cdots(E_{k_{n-1}}-E_{k_n})}\,.
\end{equation}
Using eq.(\ref{pot}), this can also be rewritten as
\begin{equation}
S_{n}(p,p')=\sum_{i_{1},\ldots,{i_{n+1}}}\alpha^{i_{1}\ldots i_{n+1}}\int_{k_1,\ldots,k_n}\frac{1}{|p-k_{1}|^{2i_{1}}\cdots |k_{n}-p^{'}|^{2i_{n+1}}}+\ldots\,,
\end{equation}
where the ellipsis denotes both quantum and super-classical contributions around $D=4$ space-time dimensions, while the 
$\alpha^{i_{1}\ldots i_{n+1}}$ are the combinations of constants that can be taken out of the integrals by properly expanding the numerators from each of the potential terms. Let us denote the classical part of this series by $S^{cl.}_{n}$
\begin{equation}
\label{sunset}
S_{n}^{cl.}(p,p')\!=\!\!\!\!\!\!\sum_{i_{1},\ldots,{i_{n+1}}}\!\!\!\!\!\alpha^{i_{1}\ldots i_{n+1}}G^{(n+1)}_{i_{i}\ldots i_{n+1}}\!(p,p')\, ,\!\!\!\! \quad G^{(n+1)}_{i_{i}\ldots i_{n+1}}\!(p,p') \equiv\! \!\int_{k_1,\ldots,k_n}\!\!\frac{1}{|p\!-\!k_{1}|^{2i_{1}}\cdots |k_{n}\!-\! p^{'}|^{2i_{n+1}}}\,.
\end{equation}
If we perform a shift in $k_{i} \rightarrow k_{i}+p \: , \:  \forall k_{i}$ in the right-hand side of eq.(\ref{sunset}) we immediately recognize the definition of a generalized sunset loop-diagram associated with a massless particle with momentum $q$ and arbitrary powers in the denominators, 
 \begin{equation}
 G^{(n+1)}_{i_{i}\ldots i_{n+1}}(q) \equiv \int_{k_1,\ldots,k_n}\frac{1}{|k_{1}|^{2i_{1}}\cdots |k_{n}-q|^{2i_{n+1}}}\,.
 \end{equation}
These integrals can be easily computed and they do share a nice factorization property in position space. In fact, by taking the Fourier transform in position space we get, using the same notation as for the one-loop case,
\begin{eqnarray}
 g^{(n+1)}_{i_{i}\ldots i_{n+1}}(r) &=& \int_{q,k_1,\ldots,k_n}\frac{e^{i q \cdot r}}{|k_{1}|^{2i_{1}}\cdots |k_{n}-q|^{2i_{n+1}}}
=\int_{q,k_1,\ldots,k_n}\frac{e^{i q \cdot r}e^{ik_{1} \cdot r}\cdots e^{i k_{n} \cdot r}}{|k_{1}|^{2i_{1}}\cdots |q|^{2i_{n+1}}}\\ &=&\bigg(\int_{k_{1}}\frac{e^{i k_{1} \cdot r}}{k_{1}^{2i_{n}}}\bigg)\cdots\bigg(\int_{k_{n}}\frac{e^{i k_{n} \cdot r}}{k_{n}^{2i_{n}}}\bigg)\bigg(\int_{q}\frac{e^{i q \cdot r}}{q^{2i_{n+1}}}\bigg)\nonumber
=g_{i_1}(r)\cdots g_{i_{n+1}}(r) ~,
\end{eqnarray}
thus generalizing to all orders what was already seen at one-loop order in the main text. 
\end{appendix}

%

\end{document}